\documentclass[aps,prd,twocolumn,showpacs,amsfonts]{revtex4}
\usepackage{amsmath}
\usepackage{graphicx}

\newcommand{\mb}[1]{\mbox{\boldmath $#1$}}
\begin{document}

\title{\bf Dynamics of radiating braneworlds}

\author{Emily Leeper, Roy Maartens and Carlos F. Sopuerta}

\affiliation{\vspace*{0.2cm} Institute of Cosmology \&
Gravitation, University of Portsmouth, Portsmouth~PO1~2EG, UK
\vspace*{0.2cm}}

\date{\today}

\begin{abstract}

If the observable universe is a braneworld of Randall-Sundrum~2
type, then particle interactions at high energies will produce
5-dimensional gravitons that escape into the bulk. As a result,
the Weyl energy density on the brane does not behave like
radiation in the early universe, but does so only later, in the
low energy regime. Recently a simple model was proposed to
describe this modification of the Randall-Sundrum~2 cosmology. We
investigate the dynamics of this model, and find the exact
solution of the field equations. We use a dynamical systems
approach to analyze global features of the phase space of
solutions.

\end{abstract}


\maketitle

\section{Introduction}

String and M theory indicate that gravity may be a
higher-dimensional interaction, while matter fields and gauge
interactions are confined to a 1+3-brane moving in the
higher-dimensional ``bulk" spacetime. These ideas have led to
relatively simple phenomenological models that allow one to
explore some of the consequences of braneworld gravity. The second
Randall-Sundrum (RS2) scenario is a five-dimensional Anti-de
Sitter (AdS$_5$) spacetime with an embedded Minkowski brane where
matter fields are confined and Newtonian gravity is effectively
reproduced at low energies. The RS2 scenario is readily
generalized to a Friedmann-Robertson-Walker (FRW) brane, and the
bulk is Schwarzschild-AdS$_5$. The modified Friedmann equation
(for a flat universe without cosmological constant) is
\begin{equation}\label{mfe}
H^2={\kappa^2 \over 3} \rho\left( 1+{\rho \over 2\lambda}
\right)+{C \over a^4}\,,
\end{equation}
where $\lambda$ is the brane tension and $C$ is the mass parameter
of the bulk black hole. The tidal field of this black hole on the
brane is felt as an effective radiative energy density, the
so-called ``dark radiation".

If the reheating and early radiation eras occur at sufficiently
high energies, this simple picture is modified. High energy
($\rho\gg \lambda \gtrsim (1~{\rm TeV})^4$) particle interactions
can produce 5D gravitons which are emitted into the bulk. This
process will contribute to an increase in the dark radiation term,
so that $C$ will not be a constant in the early universe, but will
grow until the energy scale drops below the threshold for
5D-graviton production. In the low-energy universe, $C$ will have
its asymptotic constant value, which is constrained by
nucleosynthesis limits~\cite{nucleo}.

In~\cite{Langlois:2002ke}, a simple model is constructed in which
all the 5D gravitons are assumed to be emitted radially, and their
backreaction on the bulk metric is assumed negligible, so that the
bulk is a Vaidya-AdS$_5$ spacetime. (Earlier alternative models
are developed in~\cite{hm}.) Subsequently, the corrections to this
model arising from non-radial emission have been
investigated~\cite{Langlois:2002zb}. These corrections are very
complicated, and we will confine ourselves to the simple radial
model. We find the exact solution of the dynamical equations, and
analyze the asymptotic behaviour and the global phase space of
solutions.

\section{Field equations}

The 5D field equations are
\begin{equation}\label{5}
^{(5)}G_{AB}=-\Lambda_5\, ^{(5)}g_{AB} +
\kappa_5^2\,^{(5)}T_{AB}\,, ~~\Lambda_5=-{6\over \ell^2}\,.
\end{equation}
The projected field equations on the
brane~\cite{Shiromizu:1999wj,Mizuno:2003} have three new terms
from extra-dimensional gravity:
\begin{equation}\label{modfe}
G_{\mu\nu}=-\Lambda g_{\mu\nu}+ \kappa^2 T_{\mu\nu} +
6{{\kappa^2}\over\lambda}{\cal S}_{\mu\nu} - {\cal E}_{\mu\nu}+
{\cal F}_{\mu\nu}\,,
\end{equation}
where $\Lambda = {\Lambda_5 /2} + \lambda^2 \kappa_5^4/12$ and
$\kappa^2= {\lambda }\kappa_5^4/6$. The term ${\cal S}_{\mu\nu}$
is quadratic in $T_{\mu\nu}$ and dominates at high energies
($\rho>\lambda$); this term gives rise to the $\rho^2/\lambda$
term in the modified Friedmann equation~(\ref{mfe}). The
five-dimensional Weyl tensor is felt on the brane via its
projection, ${\cal E}_{\mu\nu}$. This Weyl term is determined by
the bulk metric, not by equations on the brane, and it gives rise
to the $C/a^4$ term in the modified Friedmann
equation~(\ref{mfe}).

The term ${\cal F}_{\mu\nu}$ is also determined by the bulk
metric, and arises from $^{(5)}T_{AB}$, i.e., from 5D sources in
the bulk other than the vacuum energy $\Lambda_5$, such as a bulk
dilaton field. The presence of such sources modifies the
energy-momentum conservation equations on the
brane~\cite{Mizuno:2003}
\begin{equation}\label{modc}
\nabla^\nu T_{\mu\nu} =-2\,^{(5)}T_{AB}n^Ag^B{}_\mu\,,
\end{equation}
where $n^A$ is the unit normal to the brane.

Radial radiation in the bulk is described by
\begin{equation}\label{nrad}
^{(5)}T_{AB}= \psi k_Ak_B\,,
\end{equation}
where $k^A$ is a radial null vector and $\psi$ determines the
energy density. A physical normalization of $k^A$ is $k^Au_A=1$,
where $u^A$ is the preferred 4-velocity, coinciding on the brane
with the 4-velocity of comoving observers. Then Eq.~(\ref{nrad})
determines the form of ${\cal F}_{\mu\nu}$ as
\begin{equation}
{\cal F}_{\mu\nu}={4 \over \ell\lambda}\psi(g_{\mu\nu}+u_\mu
u_\nu)\,.
\end{equation}
When the null radiation describes 5D gravitons produced in
particle interactions in the radiation era (or in reheating with
radiative equation of state), then~\cite{Langlois:2002ke}
\begin{equation}\label{psi}
\psi= {\kappa_5^2\over 12}\, \alpha \rho^2 \,,
\end{equation}
where $\alpha$ is a dimensionless constant. If all degrees of
freedom in the standard model are relativistic, then $\alpha \sim
0.02$. The brane energy-momentum tensor on the brane is assumed to
be of the perfect-fluid type with respect to $u^A$, i.e.
$T_{\mu\nu} = (\rho+p)u_\mu u_\nu+p g_{\mu\nu}$.

For a FRW brane (which we assume to be spatially flat), the bulk
metric has 4-dimensional plane symmetry, and then Eq.~(\ref{5})
leads to the 5-dimensional Vaidya-AdS$_5$ metric~\cite{ckn}
\begin{eqnarray}
^{(5)}ds^2 &=& -F(v,r)dv^2+2drdv +r^2d\vec{x}^2\,,\\ F(v,r) &=&
{r^2\over \ell^2}-{C(v) \over r^2}\,.
\end{eqnarray}
When $C$ is independent of $v$, the metric reduces to the
Schwarzschild-AdS$_5$ metric. The coordinate $v$ is a null
coordinate (corresponding to ingoing radial light rays). The brane
trajectory is given by $r=a(t)$ and $v=v(t)$, where $t$ is proper
time for cosmic fluid particles, which move with 4-velocity
$u^\mu$.  The normalization of $u^\mu$ implies
\begin{equation}
\dot{v} = \frac{1}{F(v,a)}\left[\dot{a}+ \sqrt{\dot{a}^2 +
F(v,a)}\right]\, .
\end{equation}
For an expanding (outgoing) brane this is infinite at the apparent
horizon ($F(v,r)=0$), so we assume that an expanding brane is
always located in the region $F(v,r)>0$ (for Schwarzschild-AdS$_5$
this means outside the horizon).  See \cite{MinSasaki} for further
discussion.

The above equations lead to the modified Friedmann
equation~(\ref{mfe}), and an equation governing the evolution of
$C$~\cite{Langlois:2002ke} (valid outside the apparent horizon)
\begin{equation}\label{cd}
\dot{C} ={\kappa^2\over3}\,\alpha a^4\rho\left[
{\kappa_5^2\over6}\rho \left(1+{\rho \over \lambda}\right)-H{\rho
\over \lambda} \right].
\end{equation}
The emission of 5D gravitons is at the expense of radiation energy
density; by Eqs.~(\ref{modc})--(\ref{psi}),
\begin{equation}\label{mode}
\dot{\rho}+4H\rho=-{\kappa_5^2\over 6}\, \alpha \rho^2\,.
\end{equation}

\section{Exact solution}\label{radiating}

We introduce the dimensionless variables~\cite{Langlois:2002ke}:
\begin{equation}
\hat t={t\over \ell}\,,~ \hat H=\ell H\,,~ \hat\rho={\rho\over
\lambda}\,,~ \frac{\hat C}{a^4}= \frac{\ell^2 C}{a^4}\,.
\end{equation}
(Note that $\ell^2=6/\kappa^2\lambda$ and $\kappa_5^2=
\kappa^2\ell$.) Then the high energy regime is given by $\hat\rho,
\hat H>1$, and the dynamical equations~(\ref{mfe}), (\ref{cd}) and
(\ref{mode}) become
\begin{eqnarray}
\hat{H}^2 & = & 2\hat\rho + \hat{\rho}^2 +\frac{\hat{C}}{a^4}\,,
 \label{friedmann}\\
{d\hat\rho \over d\hat t} & = & -4\hat{H} \hat\rho -\alpha
\hat{\rho}^2\,, \label{rhodot} \\
{d\hat C \over d\hat t} & = & 2\alpha a^4\hat{\rho}^2(1+\hat \rho-
\hat H)\,.
\end{eqnarray}

These equations imply the modified Raychaudhuri equation
\begin{equation}
{d\hat H\over d\hat t} = -2\hat{H}^2-(\alpha+2)\hat{\rho}^2\,,
\end{equation}
which we use to decouple the system:
\begin{equation}
{d\over d\hat t}(\hat\rho- \hat H) = 2(\hat\rho- \hat H)^2\,.
\end{equation}
This can be integrated to give
\begin{equation}
\hat\rho -\hat H = -\frac{1}{2(\hat t-\hat{t}_0)}\,, \label{keyeq}
\end{equation}
where $\hat{t}_0$ is an integration constant.  Introducing this
into Eq.~(\ref{rhodot}), we find
\begin{equation}
{d\hat\rho \over d\hat t} = -\frac{2\hat\rho}{(\hat t-\hat{t}_0)}
- (\alpha+4)\hat{\rho}^2\,.
\end{equation}
This equation is of Bernoulli type, with solution (restoring
units)
\begin{equation}\label{rho}
\kappa^2\rho = \frac{6}{A(t-{t}_0)(t-{t}_1)}\,,
\end{equation}
where $A$ is a dimensionless integration constant and
\begin{eqnarray}
{t}_1 = {t}_0 +\ell\frac{(4+\alpha)}{A}\,.
\end{eqnarray}
Then, by Eq.~(\ref{keyeq}) the Hubble parameter can be written as
\begin{eqnarray}
H & = & \frac{1}{\alpha+4}\left\{ \frac{1}{t-
{t}_1}+\frac{2+\alpha} {2(t-{t}_0)}\right\}\,, \label{hubble}
\end{eqnarray}
and then the scale factor follows as
\begin{eqnarray}\label{sf}
a = a_0(t-t_0)^{(2+\alpha)/2(4+\alpha)} (t-t_1)^{1/(4+\alpha)}\,,
\end{eqnarray}
where $a_0$ is a constant (with dimension $\ell^{1/2}$). Finally
the expression for the dark radiation coefficient $C$ is:
\begin{eqnarray}
C & = & \frac{a_0^4[A(A-8)(t-t_1)+4\ell\{A-2(4+\alpha)\}]}
{4A^2(t-t_0)^{4/(4+\alpha)}(t-t_1)^{\alpha/(4+\alpha)}}\,.
\label{expu}
\end{eqnarray}

When $A>0$ ($t_1>t_0$), the regions in which $\rho\geq 0$ are: (i)
$t-t_1\leq 0$ and $t-t_0\leq 0$, i.e.~$ -\infty<t \leq t_0$; and
(ii) $t-t_1\geq 0$ and $t-t_0\geq 0$, i.e.~$t_1 \leq t< \infty$.
When $A<0$ ($t_1<t_0$), the $\rho\geq 0$ region is: (iii)
$t-t_0\leq 0$ and $t-t_1\geq 0$, i.e.~$ t_1 \leq t \leq t_0$. By
Eq.~(\ref{hubble}) it follows that region~(i) corresponds to
contracting models that end in a big crunch at $t=t_0$.  The
allowed part (where $F(v,a)>0$) of region~(ii) corresponds to
expanding models which start at $t=t_*\geq t_1$, where $t_*$ is
the time at which the brane is just outside the apparent horizon
(or at an initial singularity).  The allowed part of region~(iii)
corresponds to models that expand from the apparent horizon
$t=t_*$, then collapse to a big crunch at $t=t_0$.

The ever-expanding models [region (ii)] are of most relevance. By
Eq.~(\ref{sf}), it follows that near the singularity
($t\rightarrow t_1$), which corresponds to the high-energy limit,
\begin{equation}\label{he}
a(t) \sim (t-t_1)^{1/(4+\alpha)} \, ,
\end{equation}
(we note that the solution is only valid near this limit if $C$ is
initially small). This behaviour is different from that in the RS
model, $a(t)\sim t^{1/4}$, and the general relativistic one,
$a(t)\sim t^{1/2}$. Whereas $\rho$ and $H$ are always positive,
the sign of $C$ depends on the initial data, that is on the
constant $A$.  If $0<A<8$ then $C$ is always negative. If
$8<A<8+2\alpha$, then $C$ is initially negative and changes sign
at $t=t_1+4\ell [2(\alpha+4)-A]/A(A-8)$, becoming positive
forever. Finally, if $A\geq 8+2\alpha$, then $C$ is always
positive. Normally one requires $C\geq 0$ to avoid a naked
singularity in the bulk.

\begin{figure}
\begin{center}
\includegraphics[height=3in,width=2.5in,angle=-90]{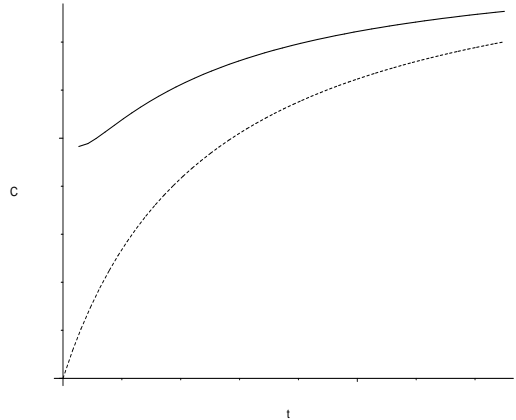}
\caption{Behaviour of the dark radiation coefficient $C$ as a
function of $t-t_1$, in the general case $A>8+2\alpha$ (solid
curve) and the special case $A=8+2\alpha$ (dashed curve). }
\label{c}
\end{center}
\end{figure}

An interesting feature is the behaviour of the dark radiation
$U=C/a^4$ near the initial time $t_*$.  In general it has a
minimum point at $t_*$ but grows quickly [this can be seen from
Eqs. (\ref{friedmann}) and (\ref{cd})].  For $C>0$ (i.e.
$A>8+2\alpha$) the general case describes the growth of the black
hole mass from a minimum initial state, when the brane is at the
horizon.  The solid line in Fig. \ref{c} illustrates this growth.

However, for the special case $A=2(4+\alpha)$ the behaviour is
completely different:
\begin{equation}
U \underset{t\rightarrow t_1}{\longrightarrow}~ {\alpha  \over
\ell^2(4+ \alpha)}\,.
\end{equation}
This special case is the one identified by perturbative analysis
in Ref.~\cite{Langlois:2002ke}, but the general case was not
given, and nor were the exact general solutions, Eqs.~(\ref{rho}),
(\ref{sf}) and (\ref{expu}).

The special case corresponds to $C=0$ at $a=0$, i.e., the black
hole is created from nothing. However, the creation begins from a
singularity ($\rho(t_1)=\infty$), so it is not clear what this
means. The growth of $C$ in this special case is illustrated by
the dotted line in Fig. \ref{c}.

In the low energy limit, the standard RS2 behaviour is recovered,
as expected:
\begin{eqnarray}
&& \rho \underset {t\rightarrow\infty} {\longrightarrow}
~\frac{6}{\kappa^2 At^2}\,,~~~ H \underset {t\rightarrow\infty}
{\longrightarrow} ~\frac{1}{2t}\,,\nonumber\\&& C \underset
{t\rightarrow\infty} {\longrightarrow}~ {a_0^4(A-8) \over 4 A}\,.
\end{eqnarray}

\section{Dynamical systems analysis}

Following~\cite{Campos:2001pa}, we use a dynamical systems
approach to investigate the global properties of the phase space
of solutions. Defining
\begin{equation}
\Omega_\rho =  \frac{2\hat\rho}{\hat{H}^2}\,,~~~ \Omega_U =
\frac{\hat U}{\hat{H}^2} = \frac{\hat C}{a^4\hat{H}^2}\,,~~~
\omega = \frac{\hat\rho}{\hat H}\,,
\end{equation}
the modified Friedmann equation~(\ref{friedmann}) takes the form
\begin{equation}
\Omega_\rho + \Omega_U + \omega^2 = 1\,. \label{constraint}
\end{equation}
In contrast to RS2 and general relativistic models, the phase
space determined by the variables $\mb{\Omega}=
(\Omega_\rho,\Omega_U,\omega)$ is not compact.  This is because
there are models in which $\Omega_U$ can change sign. We could
restrict the phase space by the physical argument that
$\Omega_U\geq 0$ because $C$ represents the mass of a black hole
in the bulk. In that case, in the restricted phase space, the
models in which $C$ changes sign will be represented by
trajectories that in the past cross the boundary of the restricted
phase space. In all cases, Eq.~(\ref{constraint}) implies that
$\Omega_U\leq 1$. Whenever $\Omega_U\geq 0$, we must have
$0\leq\Omega_\rho\leq 1$ and $-1\leq\omega\leq 1$.

\begin{figure}
\begin{center}
\includegraphics[height=4in,width=3.5in]{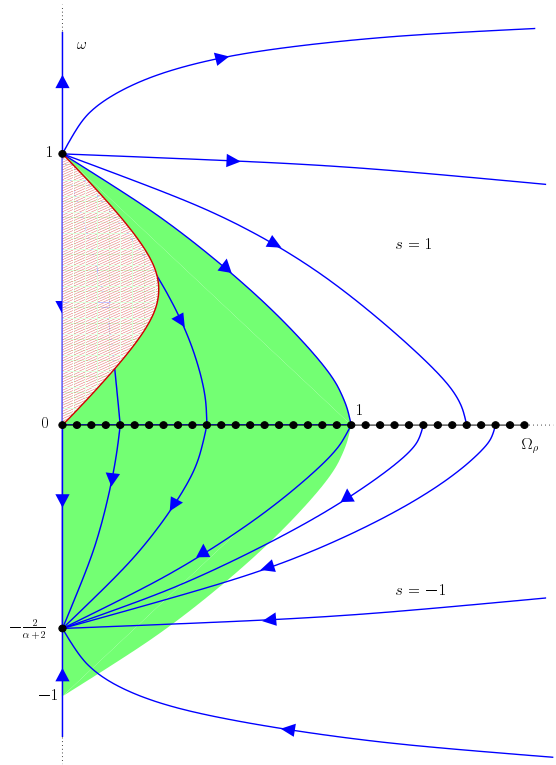}
\caption{$\Omega_\rho - \omega$ plane of the phase space. The
shaded region containing flow lines is where $C\geq0$.}
\label{orho}
\end{center}
\end{figure}

\begin{figure}
\begin{center}
\includegraphics[height=4in,width=3.5in]{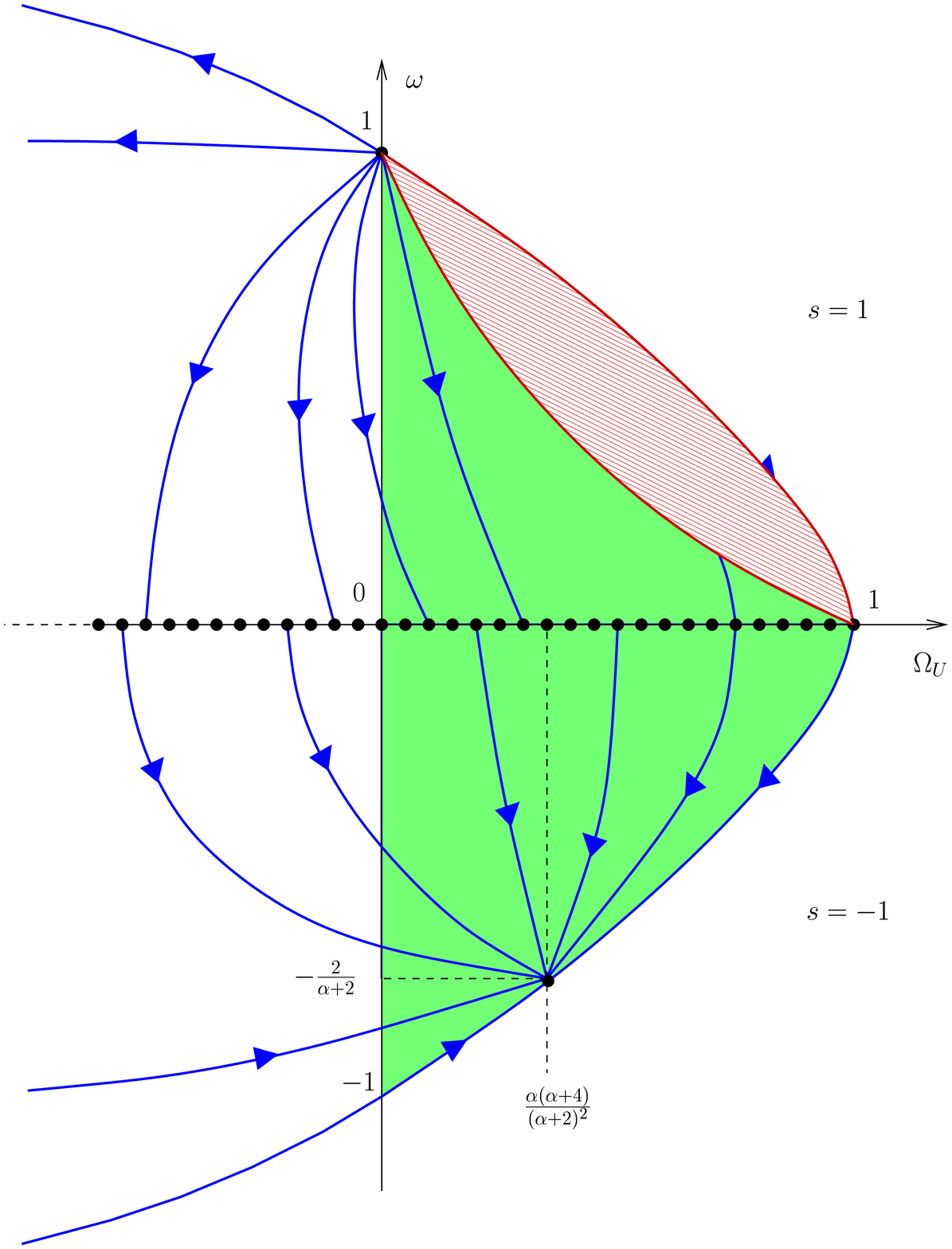}
\caption{$\Omega_U - \omega$ plane of the phase space. The shaded
region containing flow lines is where $C\geq0$.} \label{ou}
\end{center}
\end{figure}

It is convenient to rewrite the evolution equations with a new
time parameter $\tau$, where
\begin{equation}
{df \over d\tau} \equiv f' = \frac{1}{|\hat H|}\frac{df}{d\hat t}
\,. \label{newtime}
\end{equation}
Then
\begin{equation}
\hat {H}' = -s  (1+\hat q)\hat H\,, \label{dece}
\end{equation}
where $s ={\rm sgn}\,\hat H$ and $\hat q$ is the deceleration
parameter,
\begin{equation}
\hat q =  -\frac{1}{a^2\hat{H}^2}\frac{d^2{a}}{d\hat{t}^2} =
1+(\alpha+2)\omega^2\,.
\end{equation}
The evolution equations for $\mb{\Omega}$ are then
\begin{eqnarray}
\Omega_\rho' & = & s \omega\Omega_\rho
\left[2(\alpha+2)\omega-\alpha
\right]\,, \label{orhodot} \\
\Omega_U' & = & s  \omega\left[ 2(\alpha+2)\omega\Omega_U+
\alpha(\Omega_\rho+2\omega^2)-2\alpha\omega \right],
\label{oudot} \\
\omega' & = & s  \omega\left[ (\alpha+2)\omega^2-\alpha\omega-2
\right]\,. \label{omegadot}
\end{eqnarray}

Now the Friedmann equation~(\ref{constraint}) is a constraint on
the initial values of $\mb{\Omega}$. The equilibrium points of the
dynamical system $\mb{\Omega}' = \mb{f}(\mb{\Omega})$ determined
by Eqs.~(\ref{orhodot})-(\ref{omegadot}) are solutions of
$\mb{f}(\mb{\Omega}^\ast)=0$:
\begin{eqnarray}
\mb{\Omega}^\ast_1 & = & (0,0,1)\,, \\
\mb{\Omega}^\ast_2 & = & \left(0,\frac{\alpha(\alpha+4)}
{(\alpha+2)^2},-\frac{2}{\alpha+2}\right)\,, \\
\mb{\Omega}^\ast_3 & = & (\nu ,1-\nu ,0)\,.
\end{eqnarray}
Actually $\mb{\Omega}^\ast_3$ represents a set of equilibrium
points parametrized by $0\leq \nu \leq 1$.  The dynamical
character of a given equilibrium point $\mb{\Omega}^\ast$
describes the dynamical behaviour in a small neighbourhood and is
determined by the eigenvalues of the
matrix~\cite{Wainwright:1997jw} $\mb{\partial}^{}_{\mb{\Omega}}
\mb{f}$ at $\mb{\Omega}^\ast$.

The equilibrium point $\mb{\Omega}^\ast_1$ is a repeller for
expanding models. Actually, since $\omega^\ast_1 = 1 >0$, and
considering only models with positive energy density, this point
only exists for expanding models ($s  =1 ~\Leftrightarrow~H>0$).
From the fact that this equilibrium point is characterized by
$\omega=1$ it follows that $a(t)\propto t^{1/(\alpha+4)}$, and it
represents the high energy regime [see Eq.~(\ref{he})], which is
different from the one found in RS2. Moreover, at this equilibrium
point we have $\hat U=~\mbox{constant}$, so there is no flux of
gravitons.

The equilibrium point $\mb{\Omega}^\ast_2$ has $\omega^\ast_2 =
-2/(\alpha+2)<0$, so that it only exists for contracting models
($s =-1$). At this point, $a(t)\propto
(t_0-t)^{(\alpha+2)/[2(\alpha+4)]}$ and $C(t)\propto
(t_0-t)^{-6/(\alpha+4)}$. This equilibrium point, which is an
attractor, is the analogous point to $\mb{\Omega}^\ast_1$ for
expanding models. Therefore, there is an {\em asymmetry} between
expanding and contracting models, contrary to what happens in RS2
and general relativistic models~\cite{Campos:2001pa}. This is
essentially due to the fact that, unlike in RS2 and general
relativity, the dynamical equations on the brane are not invariant
under time reversal since there is energy loss to the bulk;
specifically, the energy balance equation~(\ref{mode}) is not
invariant. This asymmetry can be seen in the planes $\Omega_\rho -
\omega$ and $\Omega_U - \omega$ of the phase space shown in
Figs.~\ref{orho} and~\ref{ou}.

The set of points $\mb{\Omega}^\ast_3$ constitutes a line of
equilibrium points, which are attractors (repellers) for expanding
(contracting) models. At these points, $a(t)\propto t^{1/2}$, the
same as in standard radiation models.

\begin{figure}
\begin{center}
\includegraphics[height=3.15in, width=3.5in]{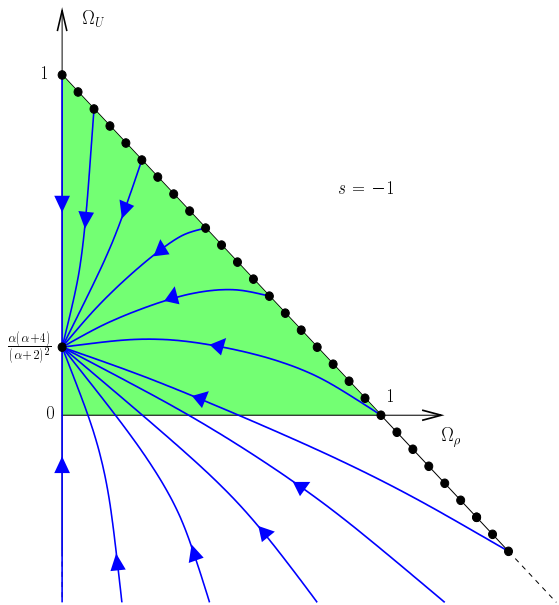}
\caption{$\Omega_U - \Omega_\rho$ plane of the phase space. The
shaded region is where $C\geq0$.} \label{ouorho}
\end{center}
\end{figure}

The phase space is a non-compact three-dimensional space. For
simplicity we only show the projections onto the three possible
2-planes: the $\Omega_\rho - \omega$ plane (Fig.~\ref{orho}), the
$\Omega_U - \omega$ plane (Fig.~\ref{ou}), and the $\Omega_U -
\Omega_\rho$ plane (Fig.~\ref{ouorho}). These planes describe
completely the dynamics of the models. In each of the figures, the
(green) shaded region containing flow lines indicates where $C$ is
non-negative. In Figs. \ref{orho} and \ref{ou} the (red) shaded
region without flow lines is the region inside the horizon where
our equations do not apply. Figures~\ref{orho} and~\ref{ou} show
the asymmetry between expanding and contracting models, and how
the dynamics is determined by the equilibrium points
$\mb{\Omega}^\ast_1$ for expanding models (repeller) and
$\mb{\Omega}^\ast_2$ for contracting models (attractor).

In Fig.~\ref{ouorho} we can see the line of equilibrium points
$\mb{\Omega}^\ast_3$ that extends to infinite values of
$\Omega_\rho$ and $\Omega_U$. (Its projection in the other two
planes can be seen in Figs.~\ref{orho} and~\ref{ou}).  This plot
describes the dynamics for contracting models only, where the
point $\mb{\Omega}^\ast_2$ appears. The corresponding one for
expanding models can be obtaining by replacing
$\mb{\Omega}^\ast_2$ by $\mb{\Omega}^\ast_1$, that is, by moving
the equilibrium point to the origin, and inverting the arrows. The
three figures show the existence of trajectories on which
$\Omega_U$, and hence $C$, changes sign, going from negative to
positive values.  Note that the contrary, to go from positive to
negative values of $C$, is forbidden. Expanding models will always
end up with a positive $C$.

\section{Conclusion}

We have investigated the dynamics of a RS2-type braneworld that is
radiating 5D gravitons into the bulk during the high-energy
radiation (or radiative reheating) era. The simplified model
developed in~\cite{Langlois:2002ke} assumes that all gravitons are
emitted radially, thus allowing the bulk spacetime to be described
by the Vaidya-AdS$_5$ metric. We found the exact solution in this
model to the field equations on the brane, as given in
Eqs.~(\ref{rho})--(\ref{expu}). The Weyl radiation term $C/a^4$ on
the brane, due to the tidal effect of the bulk black hole, is only
purely radiative (i.e., $C=\,$const) asymptotically, in the
low-energy universe. At early times, $C$ evolves due to graviton
emission, with two cases that depend on the choice of an
integration constant, as illustrated in Fig.~\ref{c}. Graviton
emission leads to a loss of energy from the brane and a consequent
asymmetry between expanding and collapsing models, that is not
present in the non-radiating RS2 braneworld. This asymmetry, as
well as the asymptotic behaviour of the models at early and late
times, is illustrated in the phase planes that fully characterize
the dynamics, shown in Figs.~\ref{orho}--\ref{ouorho}.

%%%%%%%%%%%%%%%%%%%%%%%%%%%%%%%%%%%%%%%%%%%%%%%%%%%%%%%%%%%%%%%%%%%%%%

%%%%%%%%%%%%%%%%%%%%%%%%  ACKNOWLEDGEMENTS  %%%%%%%%%%%%%%%%%%%%%%%%%%

\[ \]
{\bf Acknowledgements:} We thank Arthur Hebecker and Misao Sasaki
for useful comments. EL and CFS are supported by EPSRC. RM is
supported by PPARC.

%%%%%%%%%%%%%%%%%%%%%%%%%%%%%%%%%%%%%%%%%%%%%%%%%%%%%%%%%%%%%%%%%%%%%%

%%%%%%%%%%%%%%%%%%%%%%%%  BIBLIOGRAPHY  %%%%%%%%%%%%%%%%%%%%%%%%%%%%%%

%%%%%%%%%%%%%%%%%%%%%%%%%%%%%%%%%%%%%%%%%%%%%%%%%%%%%%%%%%%%%%%%%%%%%%


\begin{thebibliography}{99}

\bibitem{nucleo}
K. Ichiki, M. Yahiro, T. Kajino, M. Orito, and G.J. Mathews, Phys.
Rev. D{\bf 66}, 043521 (2002) [astro-ph/0203272].

\bibitem{Langlois:2002ke}
D. Langlois, L. Sorbo and M. Rodr\'{\i}guez-Mart\'{\i}nez, Phys.
Rev. Lett. {\bf 89}, 171301 (2002) [hep-th/0206146].

\bibitem{hm} A. Hebecker and J. March-Russell, Nucl. Phys. B{\bf
608}, 375 (2001) [hep-ph/0103214]; E. Kiritsis, N. Tetradis, and
T. N. Tomaras, JHEP {\bf 03}, 019 (2002) [arXiv:hep-th/0202037].

\bibitem{Langlois:2002zb}
D. Langlois and L. Sorbo, Phys. Rev. D{\bf 68}, 084006 (2003)
[hep-th/0306281].

\bibitem{Shiromizu:1999wj}
T. Shiromizu, K. Maeda, and M. Sasaki, Phys. Rev. D{\bf 62},
024012 (2000) [gr-qc/9910076].

\bibitem{Mizuno:2003}
K. Maeda, S. Mizuno, and T. Torii, Phys. Rev. D {\bf 68}, 024033
(2003) [gr-qc/0303039].

\bibitem{ckn}
A. Chamblin, A. Karch, and A. Nayeri, Phys. Lett. B{\bf 509}, 163
(2001) [hep-th/0007060].

\bibitem{MinSasaki}
M. Minamitsuji, M. Sasaki, Phys.Rev. D70 044021 (2004),
[gr-qc/0312109].


\bibitem{Campos:2001pa}
A. Campos and C.F. Sopuerta, Phys. Rev. D{\bf 63}, 104012 (2001)
[hep-th/0101060]; ibid., D{\bf 64}, 104011 (2001)
[hep-th/0105100].


\bibitem{Wainwright:1997jw}
For details on the use of dynamical systems techniques in
cosmology, see: J. Wainwright and G.F.R. Ellis, {\em Dynamical
systems in cosmology} (Cambridge University Press, 1997).

\end{thebibliography}
\end{document}